\documentclass[letterpaper]{article} 
\usepackage{aaai24}  
\usepackage{times}  
\usepackage{helvet}  
\usepackage{courier}  
\usepackage[hyphens]{url}  
\usepackage{graphicx} 
\usepackage{natbib}  
\usepackage{caption} 
\usepackage{algorithm}
\usepackage{algorithmic}

\usepackage{amssymb,amstext,amsmath}
\usepackage{multirow}
\usepackage{caption, subcaption}
\usepackage{enumitem}
\usepackage{array}
\usepackage{booktabs}
\usepackage{bbding}
\usepackage{threeparttable}
\setlist[itemize]{leftmargin=*}

\usepackage{newfloat}
\usepackage{listings}
\DeclareCaptionStyle{ruled}{labelfont=normalfont,labelsep=colon,strut=off} 
\floatstyle{ruled}
\newfloat{listing}{tb}{lst}{}
\floatname{listing}{Listing}
\nocopyright 

\title{Time-aligned Exposure-enhanced Model for Click-Through Rate Prediction}
\author{
    Hengyu Zhang\textsuperscript{\rm 1}\thanks{Work done when they were research interns at Huawei Noah’s Ark Lab.}\equalcontrib,
    Chang Meng\textsuperscript{\rm 1}\footnotemark[1]\equalcontrib,
    Wei Guo\textsuperscript{\rm 2},
    Huifeng	Guo\textsuperscript{\rm 2},
    Jieming	Zhu\textsuperscript{\rm 2},
    Guangpeng Zhao\textsuperscript{\rm 3},
    Ruiming Tang\textsuperscript{\rm 2}\footnote{Corresponding author.},
    Xiu Li\textsuperscript{\rm 1}\footnotemark[3]
}
\affiliations{
    \textsuperscript{\rm 1}Tsinghua Shenzhen International Graduate School, Tsinghua University, Shenzhen, China\\

    \textsuperscript{\rm 2}Huawei Noah's Ark Lab, Shenzhen, China\\
    \textsuperscript{\rm 3}Huawei Consumer Cloud Service Department, Shenzhen, China\\
    \{zhang-hy21, mengc21\}@mails.tsinghua.edu.cn, \{guowei67, huifeng.guo, zhaoguangpeng, tangruiming\}@huawei.com, jiemingzhu@ieee.org, li.xiu@sz.tsinghua.edu.cn
}

\begin{document}

\maketitle

\begin{abstract}

Click-Through Rate (CTR) prediction, crucial in applications like recommender systems and online advertising, involves ranking items based on the likelihood of user clicks. 
User behavior sequence modeling has marked progress in CTR prediction, which extracts users' latent interests from their historical behavior sequences to facilitate accurate CTR prediction. 
Recent research explores using implicit feedback sequences, like unclicked records, to extract diverse user interests. 
However, these methods encounter key challenges: 1) temporal misalignment due to disparate sequence time ranges and 2) the lack of fine-grained interaction among feedback sequences. 
To address these challenges, we propose a novel framework called TEM4CTR, which ensures temporal alignment among sequences while leveraging auxiliary feedback information to enhance click behavior at the item level through a representation projection mechanism. 
Moreover, this projection-based information transfer module can effectively alleviate the negative impact of irrelevant or even potentially detrimental components of the auxiliary feedback information on the learning process of click behavior. 
Comprehensive experiments on public and industrial datasets confirm the superiority and effectiveness of TEM4CTR, showcasing the significance of temporal alignment in multi-feedback modeling. 
\vspace{-3mm}

\end{abstract}

\section{Introduction}
\label{sec:intro}
With the proliferation of Internet services, individuals grapple with information overload as they struggle to seek pertinent and captivating content, underscoring the critical role of recommender systems in alleviating this challenge.
Among various recommendation techniques, Click-Through Rate (CTR) prediction within online advertising stands as paramount due to its pivotal role in revenue generation.

In the early stages of CTR prediction research, the primary focus was on improving performance through feature engineering \cite{poly2,ftrl,fm,libfm}. 
In recent years, with the advancements of deep learning techniques, CTR prediction methods based on deep learning have gained prominence \cite{xdeepfm,cheng2016wide,guo_deepfm,dcn}, which adopt neural networks to capture intricate feature interactions. 
One of the noteworthy milestones in this field is user behavior sequence modeling \cite{sasrec,gru4rec,din,dien}, employing diverse techniques including pooling techniques, RNN-based models \cite{gru4rec,lstm}, CNN-based models \cite{caser}, and self-attention models \cite{transformer,sasrec,bert4rec} to encode historical behaviors into user representations. 
Nonetheless, these early methods struggled to capture the dynamic interests of users for different target items.
DIN \cite{din} addressed this by introducing target attention to extract dynamic interests. DIEN \cite{dien} further enhanced this using a GRU network for temporal evolution modeling.

Recent advancements incorporate unclicked feedback records to enhance CTR prediction \cite{dstn,sru2b,dfn}. 
This enhancement leverages valuable insights from unclicked exposure data: it reflects users' negative interest since these items were not clicked by the users, while these exposure results are also stemming from predictions of user interest.
DSTN \cite{dstn} employs attention mechanisms for enhanced CTR prediction, while DFN \cite{dfn} fuses Wide, FM, Deep components, and Transformer architecture to capture varying-order feature interactions in explicit and implicit feedback.

\begin{figure}[]
    \centering
    \setlength{\abovecaptionskip}{2mm}
    \setlength{\belowcaptionskip}{0mm}
    \includegraphics[width=0.45\textwidth]{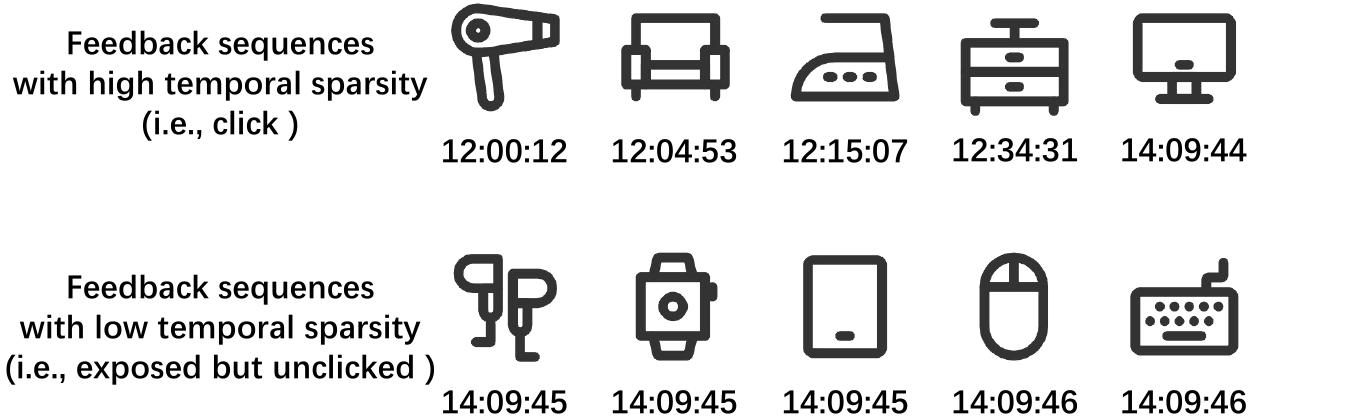}
    \caption{An example of temporal misalignment between feedback sequences with different temporal sparsity. Each item icon represents a feedback occurrence, with the associated time indicated below it.
    Due to the great differences in temporal sparsity, the feedback sequences in the figure correspond to distinctly different time ranges. }
    \label{fig:intro}
    \vspace{-5mm}
\end{figure}

Although the above CTR prediction method achieves state-of-the-art performance, they still face key challenges:

\begin{itemize}
    \item \textbf{Temporal misalignment among multi-feedback sequences.}
    Current methods \cite{dstn,dfn} truncate multi-feedback sequences (click, exposure but not click, etc.) to fixed lengths, overlooking the temporal correlations of the items among feedback sequences. 
    In fact, the temporal sparsity of each feedback varies significantly, as depicted in Figure \ref{fig:intro}.
    Some feedback may occur frequently within a second while others appear only once in a few minutes.
    Thus different fixed-length feedback sequences obtained after truncation reflect user interest across widely differing time ranges. 
    This discrepancy in time ranges limits the extent to which auxiliary feedback information can improve the prediction of the user's target behavior (i.e., click for CTR prediction).

    \item \textbf{Difficulty in performing fine-grained interactions among multi-feedback sequences.}
    Due to the aforementioned temporal misalignment problem, existing methods can only achieve sequence-level global information interaction instead of more nuanced fine-grained interaction.
    This limitation arises from the unaffordable computational cost associated with performing full interactions among items across sequences.
    However, fine-grained item-level interactions have the potential to bring greater performance improvement to the CTR prediction model.
\end{itemize}

To tackle these challenges, we propose \underline{\textbf{T}}ime-aligned \underline{\textbf{E}}xposure-enhanced \underline{\textbf{M}}odel \underline{\textbf{for}} \underline{\textbf{C}}lick-\underline{\textbf{T}}hrough \underline{\textbf{R}}ate Prediction (\textbf{TEM4CTR}) to fully leverage auxiliary exposure feedback. TEM4CTR consists of three core components: Search-based Temporal-alignment Module (STM), Projection-based Exposure-enhancement Module (PEM), and Interest Extraction Module (IEM).
STM first searches the unclicked feedback records temporally close to each click behavior as the exposure context of the corresponding click behavior.
Then it uses the attention mechanism, with the click behavior acting as the query, to extract the corresponding exposure information from the exposure context. 
This process ensures temporal alignment among the multiple feedback sequences.
PEM focuses on enhancing the representations of click behaviors by incorporating exposure information.
Guided by the correlation between the feedback sequences explicitly modeled by the projection mechanism, PEM is able to extract more valuable elements for fine-grained information enhancement.
It effectively alleviates the negative transfer phenomenon caused by the effect of useless or even harmful information in the auxiliary feedback data on the click behavior representations.
IEM employs the target attention mechanism to extract the user's latent interest from both the click sequence and its corresponding exposure information sequence.
These extracted interest representations are then concatenated to form a comprehensive user interest representation, which is used for CTR prediction.
As mentioned above, STM ensures temporal alignment among multiple feedback sequences, while PEM achieves fine-grained information enhancement on click behavior based on temporal alignment. Finally, IEM integrates the user's latent interest representations from both clicked and unclicked sequences to serve the CTR prediction task.

Our contributions can be summarized as follows:

\begin{itemize}
    \item We highlight the problem of temporal misalignment and lack of fine-grained interaction among sequences in existing methods, which may limit performance improvement.
    
    \item To address these problems, we propose a novel framework TEM4CTR to ensure temporal alignment and leverage auxiliary feedback for fine-grained information enhancement.
    
    \item We conduct comprehensive experiments on two public datasets and two industrial datasets. The experimental results demonstrate the superiority and effectiveness of our proposed TEM4CTR.
    Furthermore, the findings highlight the importance of temporal alignment in multi-feedback sequence modeling.
\end{itemize}

\section{Related Work}
In recent years, deep learning has witnessed significant advancements in various domains, such as computer vision \cite{vgg,resnet} and natural language processing \cite{lstm,ntm}. This progress has also paved the way for the development of various deep-learning-based methods for click-through rate (CTR) prediction. These deep-learning-based methods differ from traditional approaches that heavily rely on manual feature engineering, as they leverage neural networks to capture intricate feature interactions. By shifting the focus from exhaustive feature engineering to model architecture design, deep learning has propelled substantial progress in the field of CTR prediction \cite{cheng2016wide,guo_deepfm,xdeepfm,pnn}. Furthermore, recent research has been dedicated to leveraging rich historical behavioral data to capture the user's latent interest \cite{din,dien}.  For instance, DIN \cite{din} proposes a target attention mechanism to extract users' corresponding dynamic interest based on specific target items. The target attention mechanism proposed in DIN treats the target items, which are to be predicted whether users will click or not, as queries, and the user history behavior sequences as keys and values. To further improve the prediction accuracy, DIEN \cite{dien} introduces a GRU \cite{gru} network to model the temporal evolution of user interests and designs auxiliary losses to help the extraction of users' latent interest.

In recent work, there has been a focus on utilizing unclicked items in exposure sequences as auxiliary features to enhance the accuracy of click-through rate (CTR) prediction. 
DSTN \cite{dstn} incorporates the attention mechanism as an interaction layer to process the vector obtained by concatenating the target item with each feature sequence.
The output representations are then concatenated and fed into a multi-layer perceptron (MLP) for CTR prediction.
DFN \cite{dfn} introduces three components, namely Wide, FM and Deep, to model different orders of feedback interaction.
In the Deep component, Transformer \cite{transformer} is employed to model the interaction between target items and feedback sequences, while vanilla attention is utilized to capture the association among the feedback sequences.

\section{PRELIMINARIES}

\subsection{Task definition}

CTR prediction task is widely applied in recommender systems and information retrieval for online advertising. It is commonly deployed in recommendation scenarios for item ranking. CTR prediction aims to predict the probability of a user $u$ clicking on the $i$-th item $v_i$ given relevant features $\mathbf{x}_i$. Mathematically, the probability $p_i$ can be defined as:
\begin{equation}
    p_i = P(y_i=1|\mathbf{x}_i;\theta),
\end{equation}
where $y_i$ represents the label indicating whether the user clicked on the item, and $\theta$ denotes the trainable parameters of the CTR prediction. 

In order to achieve the purpose of predicting the click probability, CTR prediction models are typically trained by a binary classification problem. CTR prediction models are trained in the form of minimizing the cross-entropy loss as follows:
\begin{equation}
    L=-\frac{1}{N}\sum^N_{i=1}\left(y_ilog(p_i)+(1-y_i)log(1-p_i)\right),
\end{equation}
where $N$ is the number of interactions to be predicted.

\section{Model}
\label{model}

\begin{figure}[!ht]
    \centering
    \setlength{\abovecaptionskip}{2mm}
    \setlength{\belowcaptionskip}{0mm}
    \includegraphics[width=\linewidth]{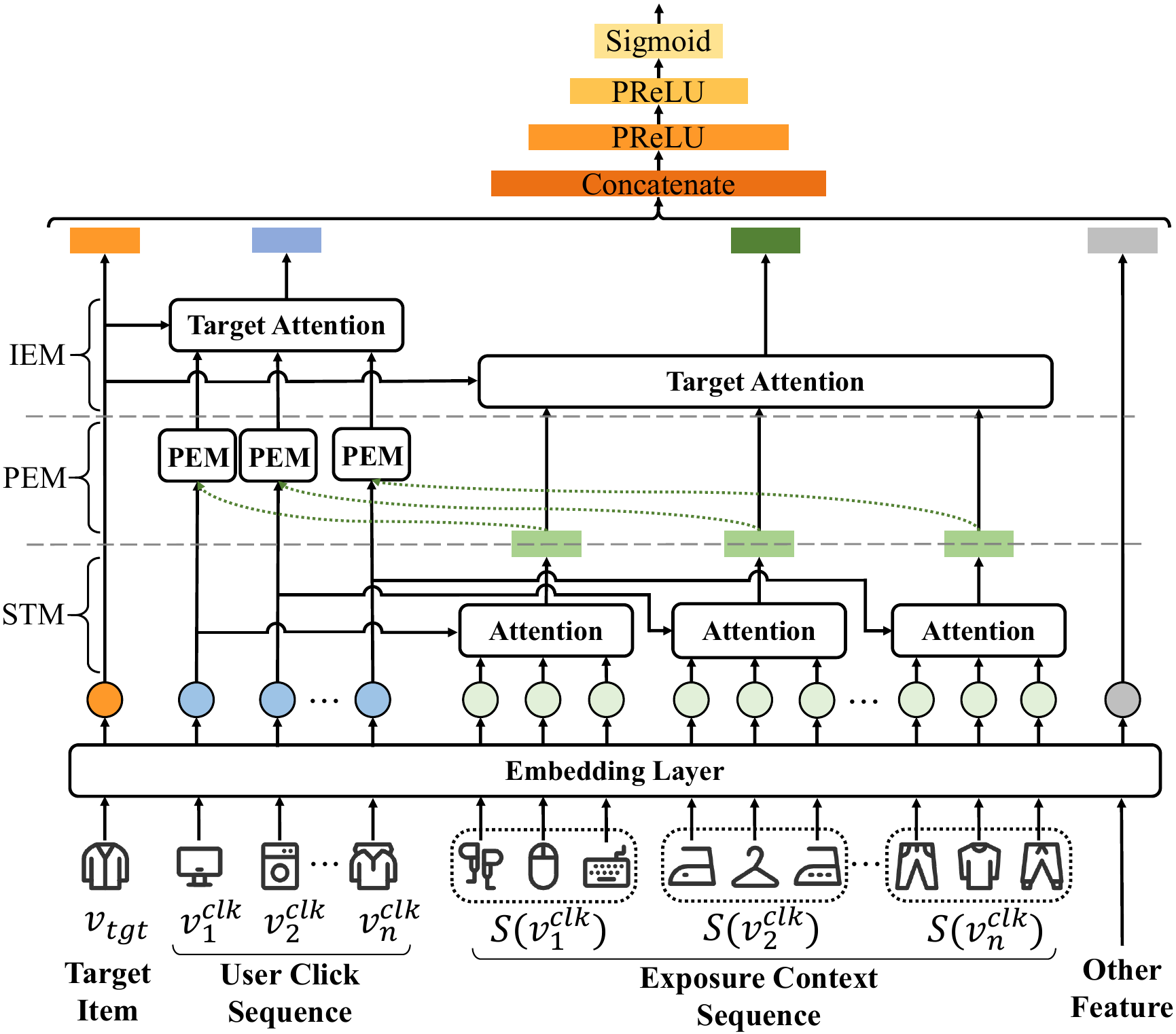}
    \caption{Overall architecture of the proposed TEM4CTR. $v^{clk}_j$ is the $j$-th item clicked by the user, and $\mathcal{S}(v^{clk}_j)$ is the exposure context corresponding to the clicked action $v^{clk}_j$. TEM4CTR ensures temporal alignment among feedback sequences, while performing fine-grained item-level information transfer between $v^{clk}_j$ and $\mathcal{S}(v^{clk}_j)$. PEM means Projection-based Exposure-enhancement Module, which utilizes the exposure context for item-level information enhancement of user click behavior representations based on the representation projection mechanism (RPM, as shown in Figure \ref{fig:rpm}).}
    \label{fig:tem4ctr}
    \vspace{-3mm}
\end{figure}

In this section, we illustrate the details of our proposed Time-aligned Exposure-enhanced Model for Click-Through Rate
Prediction (TEM4CTR).
TEM4CTR is shown in Figure \ref{fig:tem4ctr}.
Except for the embedding layer, TEM4CTR mainly consists of three components: Search-based Temporal-alignment Module (STM), Projection-based Exposure-enhancement Module (PEM), and Interest Extraction Module (IEM).
STM ensures temporal alignment among multiple feedback sequences; PEM exploits the exposure context information to perform fine-grained enhancement on the user's click behavior representations via the representation projection mechanism; IEM extracts the user's latent interest representations from multiple feedback sequences.

\subsection{Embedding Layer}
For a sequence of items $\mathcal{S}=\{v_1,v_2, v_3, \cdots, v_n\}$ displayed to a given user, the role of the embedding layer is to transform  each item $v_j$ into the $d$-dimensional embedding vector $\mathbf{e}_j$, thus obtaining the sequence of embedding vectors $\{\mathbf{e}_1,\mathbf{e}_2,\mathbf{e}_3,\cdots,\mathbf{e}_n\}$.

The datasets we use involve two types of features, one is category features and the other is high-dimensional representations extracted from the cover picture using a pre-trained network in micro-video scenarios.

The category features can be represented by one-hot vectors. Since there are thousands of users and items in the real recommendation scenario, the high-dimensional one-hot vector $\mathbf{o}^c_j$ is typically encoded as a dense embedding vector $\mathbf{e}^c_j$ by linear embedding in order to reduce the feature vector dimensionality, as follows:
\begin{equation}
    \mathbf{e}^c_j = \mathbf{E}_c\mathbf{o}^c_j,
\end{equation}
where $\mathbf{E}_c$ is the trainable embedding matrix.

In the Microvideo-1.7M dataset, $\mathbf{o}^{f}_{j} \in \mathbb{R}^{d_f}$ is the high-dimensional feature vectors in the micro video scene extracted from the cover images by Inception-v3 pre-trained on ImageNet \cite{thacil}, and ${d_f}$ is the dimension of $\mathbf{e}^c_j$. To reduce the complexity of the model, we use linear embedding to project them into the $d$-dimensional space according to the following equation: 
\begin{equation}
    \mathbf{e}^f_j = \mathbf{E}_f\mathbf{o}^f_j,
\end{equation}
where $\mathbf{E}_f\in\mathbb{R}^{d\times d_f}$ is the trainable linear mapping embedding matrix, and $d$ is the embedding dimensionality.

Lastly, we concatenate all the obtained embedding vectors together to get the final embedding vector $\mathbf{e}_j$.

Depending on whether the user has clicked with the item, the clicked subsequence $\mathcal{S}_{clk}=\{v^{clk}_1, v^{clk}_2,\cdots\}$ and unclicked subsequence $\mathcal{S}_{unclk}=\{v^{unclk}_1, v^{unclk}_2,\cdots\}$ can be obtained from the sequence $\mathcal{S}$, respectively.
The corresponding sets of embedding vectors for the clicked subsequence $\mathcal{S}_{clk}$ and the unclicked subsequence $\mathcal{S}_{unclk}$ are $\mathcal{S}^{emb}_{clk}$ and $\mathcal{S}^{emb}_{unclk}$, respectively.

\subsection{Search-based Temporal-alignment Module (STM)}

STM aims to ensure the temporal alignment among multi-feedback sequences, while extracting the exposure context information corresponding to each click behavior for subsequent fine-grained information enhancement in PEM.

\subsubsection{Exposure Context Search}

To assure temporal alignment between clicked and unclicked sequences, we search for unclicked items in exposure close to the occurrence time of the clicked behavior as the exposure background context of the clicked behavior.

\begin{figure}[]
    \centering
    \setlength{\abovecaptionskip}{1mm}
    \setlength{\belowcaptionskip}{0mm}
    \includegraphics[width=0.38\textwidth]{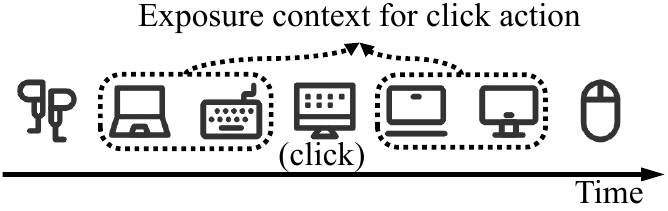}
    \caption{A Example of Exposure Context Search ($l=4$). }
    \label{fig:ecs}
    \vspace{-3mm}
\end{figure}

For the clicked item $v^{clk}_j$ in the clicked sequence $\mathcal{S}_{clk}$, we search for $l$ unclicked items before and after the occurrence of the click behavior according to the timestamp, as shown in Figure \ref{fig:ecs}.
Thus we can obtain the set $\mathcal{S}(v^{clk}_j)$ of unclicked items as the exposure background context corresponding to the click behavior $v^{clk}_j$.
And its corresponding set of embedding vectors is $\mathcal{S}(v^{clk}_j)^{emb}$.

\subsubsection{Exposure Information Extraction}

We apply the attention mechanism to sufficiently extract the exposure contextual information. 
Attention mechanism contains three main elements: query ($\mathbf{q}$), key ($\mathbf{k}$) and value ($\mathbf{v}$).
The attention mechanism can be formulated as the following equation:
\begin{equation}
    f(\mathbf{q}, \mathbf{k}) = [\mathbf{q}, \mathbf{k}, \mathbf{q}-\mathbf{k}, \mathbf{q}\cdot \mathbf{k}],
\end{equation}
\begin{equation}
    score(\mathbf{q}, \mathbf{k}) = softmax\left(\text{MLP}\left((f(\mathbf{q}, \mathbf{k})\right)\right),
\end{equation}
\begin{equation}
    \text{Attention}(\mathbf{q}, \mathbf{k}, \mathbf{v}) = score(\mathbf{q}, \mathbf{k}) \cdot \mathbf{v},
\end{equation}
where $(\cdot)$ means Hadamard product, i.e. element-wise product, and MLP is a simple three-layer multilayer perceptron.

In the STM module, clicked item $v^{clk}_j$ acts as \textit{query} and its corresponding sets $\mathcal{S}(v^{clk}_j)$ of unclicked items in exposure act as \textit{key} and \textit{value}.
The attention mechanism can generate the weights of each item in the unclicked sets based on the relationship between clicked and unclicked items.
\begin{equation}
f(\mathbf{e}^{clk}_j, \mathbf{e}^{unclk}_k) = [\mathbf{e}^{clk}_j, \mathbf{e}^{unclk}_k, \mathbf{e}^{clk}_j-\mathbf{e}^{unclk}_k, \mathbf{e}^{clk}_j\cdot \mathbf{e}^{unclk}_k], 
\end{equation}
\begin{equation}
    \begin{aligned}
    score(\mathbf{e}^{clk}_j, \mathbf{e}^{unclk}_k) = softmax\left(\text{MLP}\left((f(\mathbf{e}^{clk}_j, \mathbf{e}^{unclk}_k)\right)\right),\\\mathbf{e}^{unclk}_k \in \mathcal{S}(v^{clk}_j)^{emb}.
    \end{aligned}
\end{equation}

Thus, we can extract the exposure contextual information $\mathbf{c}_j$ of the click behavior $v^{clk}_j$.
\begin{equation}
    \mathbf{c}_j = \sum_{\mathbf{e}^{unclk}_k \in \mathcal{S}(v^{clk}_j)^{emb}} score(\mathbf{e}^{clk}_j, \mathbf{e}^{unclk}_k) \mathbf{e}^{unclk}_k.
\end{equation}

\subsection{Projection-based Exposure-enhancement Module (PEM)}

As analyzed in the Introduction Section, under the premise that STM ensures temporal alignment, fine-grained item-level information interaction is available.
In order to extract useful knowledge from the exposure context information to perform fine-grained enhancement on the click behavior information, we use a representation projection mechanism \cite{dumn,hpmr} to model the correlation between the click information and the exposure context information.
The correlation-guided information transfer alleviates the negative effect led by useless or even harmful contents of the exposure context information on the click behavior representations.

\begin{figure}[!ht]
    \centering
    \setlength{\abovecaptionskip}{1mm}
    \setlength{\belowcaptionskip}{0mm}
    \includegraphics[width=0.25\textwidth]{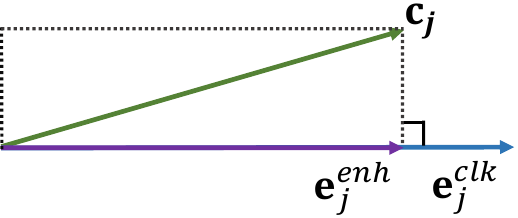}
    \caption{Illustration of Representation Projection Mechanism (RPM).}
    \label{fig:rpm}
    \vspace{-3mm}
\end{figure}

Based on the representation projection mechanism as shown in Figure \ref{fig:rpm}, we can project the exposure context representation onto the click behavior representation to obtain the enhanced information $\mathbf{e}^{enh}_j$:
\begin{equation}
\mathbf{e}^{enh}_j=\frac{\langle\mathbf{e}^{clk}_j,\mathbf{c}_j\rangle}{|\mathbf{e}^{clk}_j|}\frac{\mathbf{e}^{clk}_j}{|\mathbf{e}^{clk}_j|},
\end{equation}
where $\langle \mathbf{a},\mathbf{b}\rangle$ means the vector inner product of $\mathbf{a}$ and $\mathbf{b}$, and $|\mathbf{a}|$ is the L2 norm of $\mathbf{a}$.

Then, we add the enhanced information to the click behavior representation for representation enhancement, and obtain the enhanced click behavior representation $\mathbf{e}^{clk*}_j$ as follows:
\begin{equation}
    \mathbf{e}^{clk*}_j=\mathbf{e}^{clk}_j+\mathbf{e}^{enh}_j.
\end{equation}

\subsection{Interest Extraction Module (IEM)}

IEM is a replaceable component that can adopt most of the mainstream CTR models to extract user interest representations, and we use DIN \cite{din} by default.

Following DIN \cite{din}, we adopt the target attention mechanism to extract interest in enhanced clicked  
representation $\mathbf{e}^{clk*}_j$ and exposure contextual information $\mathbf{c}_j$, respectively.

Click interest representation $\mathbf{h}$ extracted from enhanced clicked representation $\mathbf{e}^{clk*}_j$ can be calculated by the following formula:
\begin{equation}
    score(\mathbf{e}_{tgt}, \mathbf{e}^{clk*}_j) = softmax\left(\text{MLP}\left((\mathbf{e}_{tgt}, \mathbf{e}^{clk*}_j)\right)\right),
\end{equation}
\begin{equation}
    \mathbf{h}=\sum^n_{j=1}score(\mathbf{e}_{tgt}, \mathbf{e}^{clk*}_j) \mathbf{e}^{clk*}_j,
\end{equation}
where $\mathbf{e}_{tgt}$ means the embedding vector of target item $v_{tgt}$.

In a similar way, we can obtain unclick interest representation $\mathbf{g}$ extracted from exposure contextual information $\mathbf{c}_j$ via target attention mechanism:
\begin{equation}
    score(\mathbf{e}_{tgt}, \mathbf{c}_j) = softmax\left(\text{MLP}\left((\mathbf{e}_{tgt}, \mathbf{c}_j)\right)\right),
\end{equation}
\begin{equation}
    \mathbf{g}=\sum^n_{j=1}score(\mathbf{e}_{tgt}, \mathbf{c}_j) \mathbf{c}_j.
\end{equation}

\subsection{Training Objective}

We concatenate target item embedding, clicked interest representations, and unclicked interest representations for the CTR prediction task.
Then, we predict the click probability $p$ based on the concatenated representations via MLP as follows:
\begin{equation}
    p=sigmoid(\text{MLP}([\mathbf{e}_{tgt},\mathbf{h},\mathbf{g}])).
\end{equation}

TEM4CTR is trained with cross-entropy loss, and the loss function can be calculated by:
\begin{equation}
    L=-\frac{1}{N}\sum^N_{i=1}\left(y_ilog(p_i)+(1-y_i)log(1-p_i)\right),
\end{equation}
where $N$ is the number of interaction pairs to be predicted, and $y_i$ is the label of whether or not the $i$-th interaction pair occurs a click action.

\subsection{Time Complexity Analysis}
The operation of exposure context search in the STM of TEM4CTR can be implemented by data preprocessing in an offline way, so it does not take up the model's online inference time.
So the time complexity of STM is mainly due to the exposure context extraction.
In the exposure context extraction process, STM uses the attention mechanism to output weighted representations of all retrieved unclicked items in exposure, which costs $\mathcal{O}(nld)$. Besides, the time complexity of PEM is $\mathcal{O}(nd)$, and the time complexity of the scoring process in IEM is $\mathcal{O}(nd)$.
In summary, the total complexity of TEM4CTR is $\mathcal{O}(nld)$.

\section{Experiments}

We conduct comprehensive experiments on two widely-used public datasets to explore the following research questions:
\begin{itemize}
\item {\bfseries RQ1:} How does TEM4CTR perform compared to the state-of-the-art CTR prediction models?
\item {\bfseries RQ2:} How do different modules affect the performance of TEM4CTR respectively?
\item {\bfseries RQ3:} Can TEM4CTR be compatible with mainstream CTR prediction models?
\item {\bfseries RQ4:} How does TEM4CTR perform when applied on industrial datasets?
\item {\bfseries RQ5:} How do different hyperparameters affect the performance of TEM4CTR?
\end{itemize}
Due to space limitations, hyperparameter analysis results (\textbf{RQ5}) are provided in \textbf{Appendix C} in the supplementary materials.

\subsection{Experimental Setting}
To comprehensively validate the superior performance of our proposed TEM4CTR model, we conduct experiments on two public real-world recommendation datasets. These public datasets contain both clicked and unclicked feedback data, and the details of the public datasets are as follows:
\begin{itemize}
\item \textbf{Alibaba}\footnote{https://tianchi.aliyun.com/dataset/56}: This is a Taobao click-through rate prediction dataset provided by Alibaba, collected by randomly sampling the ad display/click logs of 1.14 million users (26 million records) from the Taobao.com over an 8-day period. As described on the official website of the dataset, we used the records of the previous 7 days (2017.05.06-2017.05.12) for the training sample and the records of the 8-th day (2017.05.13) for the test sample.
\item \textbf{Microvideo-1.7M}\footnote{https://github.com/Ocxs/THACIL}: This dataset comes from real data of micro-video sharing service in China which contains 1.7 million micro-videos \cite{thacil}. Microvideo-1.7M contains over 12 million interaction records from 10,986 users in response to 1.7 million short videos. For each short video sample, the dataset contains 512-dimensional features extracted from the short video cover by the Inception-v3 model pre-trained on the ImageNet dataset. In addition, the dataset contains category ids with a total of 512 categories.

\end{itemize}

For each user, their clicked items are sorted by timestamp, and then the exposure context of each clicked behavior is searched as described in STM via data preprocessing.
We follow the official splitting settings corresponding to the dataset to obtain training and test sets. 
For the training set, we randomly sample subsequences as training samples.
Supposing there are $T$ historical behaviors for a subsequence, 1 to $T-1$ click records and the corresponding exposure contexts are used to predict the click probability of the $T-th$ target item.
For the test set, supposing there are $T$ historical behaviors for a user, 1 to $T-1$ click records and the corresponding exposure contexts are used to predict the target item $T$.
For each user, we randomly select a non-clicked item to serve as the negative sample. Table \ref{tab:dataset} shows the statistics of the two public datasets.

\begin{table}[]
\centering
\setlength{\abovecaptionskip}{2mm}
\setlength{\belowcaptionskip}{0mm}
\resizebox{\linewidth}{!}{
\begin{tabular}
{@{}c|c|c|c@{}}
\toprule
Dataset & \# Users & \# Items & \# Interactions\\ \midrule
Alibaba & 1,141,729 & 846,811 & 26,557,961  \\
Microvideo-1.7M & 10,986 & 1,704,880 & 12,737,619 \\
\bottomrule
\end{tabular}}
\caption{Dataset Statistics}
\label{tab:dataset}
\vspace{-3mm}
\end{table}

\subsubsection{Evaluation Metrics}
We adopt the widely-used evaluation metric AUC to evaluate the CTR prediction performance.
AUC measures the goodness of assigning positive samples higher scores than randomly chosen negative samples. 
A higher AUC value indicates better performance. 

We adopt the RelaImpr metric to measure the relative performance improvement of the CTR prediction model. For a random guesser, the value of the AUC is 0.5. Thus, RelaImpr is defined as follows \cite{din}:
\begin{equation}
    RelaImpr = \left(\frac{\text{AUC}-0.5}{\text{AUC}_{base}-0.5}-1\right)\times 100\%.
\label{equ:auc_rel}
\end{equation}

\subsubsection{Comparison Baselines}
To comprehensively demonstrate the effectiveness of the proposed TEM4CTR, we compared it with nine state-of-the-art recommendation baselines: AvgPool DNN, RUM \cite{rum}, RNN \cite{mimn}, GRU4Rec \cite{gru4rec}, DIN \cite{din}, DIEN \cite{dien}, DSTN \cite{dstn}, DFN \cite{dfn}, CAN \cite{can}.
The introduction to the baseline methods can be found in \textbf{Appendix B} in the supplementary materials.

For fair comparisons, all models including TEM4CTR and baselines use the same number of unclicked exposure records.

\subsubsection{Implementation Details}

We follow the TensorFlow implementations of AveragePooling DNN, RUM, ARNN, GRU4Rec, DIN, and DIEN in the open source code of MIMN \cite{mimn}. For DSTN, DFN and CAN, we followed their official implementations. 

TEM4CTR is also implemented in the Tensoflow \cite{TensorFlow} framework. 
For all methods, we set the maximum length of the click sequence to $n=30$, following the setting of DFN \cite{dfn}. We set the exposure context range $l$ of TEM4CTR to 10. Therefore, for a fair comparison, the maximum length of the exposed but unclicked sequences in the experiments of baseline models was set to $l\times n$, to ensure that they used the same amount of unclicked feedback data as TEM4CTR. For all methods, the dimensionality of the embedding vector is set to 16. We use the Adam \cite{adam} optimizer with a learning rate of 0.005 for training. And the batch size is set as 256 for both training and testing.

\subsection{RQ1: Overall Performances}

\begin{table}[]
\setlength{\abovecaptionskip}{2mm}
\setlength{\belowcaptionskip}{0mm}

\centering
\resizebox{\linewidth}{!}{
\begin{tabular}{lcccc}
\toprule
\multirow{2}{*}{Model} & \multicolumn{2}{c}{Alibaba}                             & \multicolumn{2}{c}{Microvideo-1.7M}                     \\ 
& \multicolumn{1}{c}{AUC} & \multicolumn{1}{c}{RelaImpr} & \multicolumn{1}{c}{AUC} & \multicolumn{1}{c}{RelaImpr} \\ \hline
 AvgPool DNN&  \multicolumn{1}{c}{0.8702}&0.00\% 
&  \multicolumn{1}{c}{0.8579}&0.00\% 
\\
 RUM&   \multicolumn{1}{c}{0.8722}&0.54\% 
&  \multicolumn{1}{c}{0.8647}&1.90\% 
\\
 ARNN&   \multicolumn{1}{c}{0.8726}&0.65\% 
&  \multicolumn{1}{c}{0.8666}&2.43\% 
\\
 GRU4Rec&   \multicolumn{1}{c}{0.8746}&1.19\% 
&  \multicolumn{1}{c}{0.8674}&2.65\% 
\\
 DIN&   \multicolumn{1}{c}{0.8751}&1.32\% 
&  \multicolumn{1}{c}{0.8691}&3.13\% 

\\
 DIEN&   \multicolumn{1}{c}{0.8759}&1.54\% 
&  \multicolumn{1}{c}{0.8697}&3.30\% 
\\
 DSTN&   \multicolumn{1}{c}{0.8837}&3.65\% 
&  \multicolumn{1}{c}{0.8703} &3.46\% \\
 DFN&   \multicolumn{1}{c}{0.8942} &6.48\% 
&  \multicolumn{1}{c}{0.8717} &3.86\% \\
CAN & \multicolumn{1}{c}{0.8879} &4.78\% 
&  \multicolumn{1}{c}{0.8716} &3.83\%
\\\hline
 TEM4CTR&   \multicolumn{1}{c}{\textbf{0.9393}$^*$} & \textbf{18.67\%}
&  \multicolumn{1}{c}{\textbf{0.8808}$^*$} & \textbf{6.40\%} 
\\
 \bottomrule
\end{tabular}}
\caption{Performance comparison using different methods on two public recommendation datasets. The best performance method is denoted in bold fonts. The "*" mark denotes the statistical significance (p\textless 0.05) of comparing TEM4CTR with the strongest baseline results. RelaImpr is calculated according to Equation \ref{equ:auc_rel}.}
\label{tab:comparison}
\vspace{-4mm}
\end{table}

The overall performance comparison of our proposed TEM4CTR model and baselines is shown in Table \ref{tab:comparison}.
Based on the results, we summarize the following observations:

\begin{itemize}
    \item TEM4CTR achieved the best AUC metrics on all two datasets, demonstrating the superiority and effectiveness of our proposed method.
Compared with the base model DNN, TEM4CTR achieves a relative improvement (according to Equation \ref{equ:auc_rel}) of 18.67\% and 6.40\% on the Alibaba and Microvideo-1.7M datasets, respectively.
The excellent performance of TEM4CTR shows the importance of temporal alignment in multi-feedback sequence modeling.
Under the premise of temporal alignment, fine-grained information enhancement can be performed between feedback sequences, thus better utilizing auxiliary feedback information to improve CTR prediction.
In a follow-up experiment, we further validated the superior performance of TEM4CTR on two industrial datasets.

\item Modeling the correlations between multi-feedback sequences can bring performance improvements to the CTR prediction task.
Although baselines other than DFN (such as DIN, DIE, DSTN, CAN, etc.) also utilize unclicked records as auxiliary feedback data, they all handle different sequences independently.
In contrast to them, DFN achieves some performance improvement by modeling the relationship between click and unclick sequences.
This improvement is limited because there are significant time-range differences in user interest extracted from different sequences due to the different sparsity of different feedbacks and the truncation operation used by DFN for different sequences, as analyzed in the Introduction Section.
Our proposed TEM4CTR models time-aligned multi-feedback sequence correlations, and achieves significant performance gains as a result.
It is because TEM4CTR avoids the performance degradation  caused by temporal-misalignment among multiple feedback interest.
\end{itemize}

\subsection{RQ2: Ablation Study}

\begin{table}[]
\setlength{\abovecaptionskip}{2mm}
\setlength{\belowcaptionskip}{0mm}
\centering
\begin{tabular}{ccc}
\toprule
Model & Alibaba & Microvideo-1.7M \\\hline
 TEM4CTR w/o STM &  0.8747  &  0.8730 \\
 TEM4CTR w/o PEM &  0.9326  &  0.8799 \\ 
 TEM4CTR w/o IEM &  0.9367  &  0.8759 \\\hline
 TEM4CTR &  \textbf{0.9393}  &  \textbf{0.8808}  \\ \bottomrule
\end{tabular}
\caption{Ablation study results of different consisting modules in TEM4CTR.}
\label{tab:ablation}
\vspace{-5mm}
\end{table}

As introduced in the Model Section, the TEM4CTR model consists of three main components, namely the STM, PEM and IEM modules.
To verify the effectiveness of these three components, we perform an ablation study by removing each component from TEM4CTR, which results in the following three variants:
\begin{itemize}
\item {\bfseries TEM4CTR w/o STM:} We removed the STM in TEM4CTR, i.e., broke the temporal alignment. Since the temporal alignment is not satisfied, the item-level information enhancement in PEM also needs to be adjusted to sequence-level information enhancement. 
\item {\bfseries TEM4CTR w/o PEM:} We remove the PEM from TEM4CTR, i.e., we delete the item-level information enhancement of the exposure feedback information on the click behavior.
\item {\bfseries TEM4CTR w/o IEM:} We replace the target attention mechanism adopted as the default by IEM in TEM4CTR with an average pooling layer.
\end{itemize}

The results of the comparison between TEM4CTR and the variants are shown in Table \ref{tab:ablation}. According to the results in Table \ref{tab:ablation}, we draw the following conclusions:
\begin{itemize}
    \item It is clear that removing any component will result in performance degradation, which proves that all components of our proposed TEM4CTR are superior and necessary. 
    \item We can find that TEM4CTR w/o STM performs the worst, relatively decreasing 14.71\% and 2.05\% on Alibaba and Microvideo datasets, respectively, which fully demonstrates the importance of temporal-alignment. With the help of STM, TEM4CTR can fully extract useful contextual information from unclicked feedback in exposure, helping the model tap into the user's potential interest.
    \item Compared to TEM4CTR, the performance of TEM4CTR w/o PEM decreases consistently on two datasets. This indicates that the projection mechanism has successfully extracted useful information from the context representation. With the projection mechanism, our method explicitly models the correlation between click information and exposure context information, thus further extracting useful information to perform fine-grained enhancement on click representation. Meanwhile, this alleviates the phenomenon of negative transfer led by useless or even harmful contents of the exposure context information.
    \item The performance of TEM4CTR w/o IEM is also consistently lower than that of TEM4CTR on the two datasets, which demonstrates the importance of making full use of target items to perform more accurate prediction of users' latent interest from exposure context information representations and enhanced click representations.
\end{itemize}

\subsection{RQ3: Compatibility Analysis}

TEM4CTR is not only a specific model, but also a general framework that can be applied to combine with existing CTR prediction models.
To verify our claim, we use some representative baseline models to act as IEMs in TEM4CTR and evaluate the resulting models by comparing them with the original baselines.
The selected CTR prediction models include AveragePooling DNN, DIN \cite{din}, DIEN \cite{dien} and DFN \cite{dfn}, and the comparison results are shown in Table \ref{tab:compatibility}.

\begin{table}[]
\setlength{\abovecaptionskip}{2mm}
\setlength{\belowcaptionskip}{0mm}
\centering
\begin{tabular}{lcc}
\toprule
Model               & Alibaba & Microvideo-1.7M \\ \hline
AvgPool DNN & 0.8702  & 0.8579          \\
TEM4CTR (AvgPool)   & 0.9367  & 0.8759          \\ \hline
DIN                 & 0.8751  & 0.8691          \\
TEM4CTR (DIN)       & 0.9393  & 0.8808          \\ \hline
DIEN                & 0.8759  & 0.8697          \\
TEM4CTR (DIEN)      & 0.9427  & 0.8803          \\ \hline
DFN                 & 0.8942  & 0.8717          \\
TEM4CTR (DFN)       & 0.9446  & 0.8811          \\ \bottomrule
\end{tabular}
\caption{Compatibility analysis with different baselines acting as IEM of TEM4CTR on two public datasets.}
\label{tab:compatibility}
\vspace{-5mm}
\end{table}

We can observe that, for each baseline applied to TEM4CTR, a consistent performance improvement is achieved on two public datasets. The performance improvement is mainly attributed to the advantage of temporal alignment and the effective fine-grained information enhancement on the click behavior representation from the exposure context.
Although DIEN and DFN applied in TEM4CTR achieve better performance than DIN, their complexity is relatively higher. So we choose DIN as the default option to balance performance and efficiency.

The above results demonstrate the wide generality and compatibility of TEM4CTR, which can provide significant performance improvements to existing CTR prediction models. It also demonstrates the importance of temporal alignment in multi-feedback modeling.

\subsection{RQ4: Application on Industrial Dataset}

Besides the public datasets, TEM4CTR achieves promising results in real-world industrial datasets. In this section, we present the offline evaluation results obtained in the paid-to-promote scenario of one mainstream \textit{Mobile App Store}. The app gallery has tens of millions of daily active users, which generates billions of user log events every day in feedback such as browsing, clicking, and downloading apps.

\subsubsection{Experiment Description.}

The offline evaluation is conducted over two industrial datasets collected and sampled from eight-day logs of two scenarios on the Mobile App Store. The first seven consecutive days’ log is used for training, and the eighth-day log is for testing. 
We collect user feedback data over the past month for user interest modeling, which includes user click behavior and corresponding unclicked feedback in exposure during the past month.
Due to the larger amount of industrial scene data, we simplify the TEM4CTR by replacing the exposure information extraction in STM with a pooling operation and replacing the PEM with concatenating vector.
So we validate the TEM4CTR performance by adding the exposure context information (after pooling aggregation) as a new feature to the Base Model. 
Namely, the feature of exposure context information is concatenated with other regular features and acts as the input to TEM4CTR. 
The Base Model is the actual recommendation model used on the Mobile App Store, and the specific implementation cannot be disclosed due to commercial confidentiality reasons.

\subsubsection{Performance Comparison.}
We leverage the commonly-used offline metric AUC to evaluate our TEM4CTR, where the calculation of relative improvement is consistent with Equation \ref{equ:auc_rel}. As shown in Table \ref{tab:industry}, our method is compared with the Base Model, relatively promoting \textbf{2.39\%} and \textbf{2.29\%} on Scenario 1 and Scenario 2 according to Equation \ref{equ:auc_rel}, respectively. The significant improvement of our method demonstrates that it is effective and superior in large-scale industrial scenarios and has high practical value.

\begin{table}[]
    \setlength{\abovecaptionskip}{2mm}
    \setlength{\belowcaptionskip}{0mm}
    \centering
    \begin{threeparttable}
    \resizebox{\linewidth}{!}{
    \begin{tabular}{c|cc|cc}
    \toprule
    {Dataset}&
    \multicolumn{2}{c}{Scenario 1}&
    \multicolumn{2}{|c}{Scenario 2}\cr
    \cmidrule(lr){1-1}
    \cmidrule(lr){2-3}\cmidrule(lr){4-5}
    {Metrics}&AUC&RelaImpr&AUC&RelaImpr\cr
    \midrule
    Base Model &0.8926&-&0.8879&-\cr
    \textbf{Ours} &\textbf{0.9020}&\textbf{2.39\%}&\textbf{0.8968}&\textbf{2.29\%}\cr
    \bottomrule\bottomrule
    \end{tabular}}
    \end{threeparttable}
    \caption{Performance comparison on industrial datasets. RelaImpr is calculated according to Equation \ref{equ:auc_rel}.}
    \label{tab:industry}
    \vspace{-5mm}
\end{table}

\section{CONCLUSIONS}

In this paper, we highlight the temporal misalignment problem and the lack of fine-grained interactions among sequences in existing multi-feedback sequence modeling work. To address these problems, we propose a novel framework TEM4CTR to improve the performance of CTR prediction by effectively utilizing the auxiliary feedback information under the premise of temporal alignment.
Extensive experiments on public datasets and industrial scenarios demonstrate TEM4CTR's superior effectiveness and broad compatibility.

\bibliography{ref}

\begin{thebibliography}{32}
\providecommand{\natexlab}[1]{#1}

\bibitem[{Abadi et~al.(2016)Abadi, Barham, Chen, Chen, Davis, Dean, Devin,
  Ghemawat, Irving, Isard, Kudlur, Levenberg, Monga, Moore, Murray, Steiner,
  Tucker, Vasudevan, Warden, Wicke, Yu, and Zheng}]{TensorFlow}
Abadi, M.; Barham, P.; Chen, J.; Chen, Z.; Davis, A.; Dean, J.; Devin, M.;
  Ghemawat, S.; Irving, G.; Isard, M.; Kudlur, M.; Levenberg, J.; Monga, R.;
  Moore, S.; Murray, D.~G.; Steiner, B.; Tucker, P.~A.; Vasudevan, V.; Warden,
  P.; Wicke, M.; Yu, Y.; and Zheng, X. 2016.
\newblock TensorFlow: {A} System for Large-Scale Machine Learning.
\newblock In Keeton, K.; and Roscoe, T., eds., \emph{12th {USENIX} Symposium on
  Operating Systems Design and Implementation, {OSDI} 2016, Savannah, GA, USA,
  November 2-4, 2016}, 265--283. {USENIX} Association.

\bibitem[{Bian et~al.(2022)Bian, Wu, Ren, Pi, Zhang, Xiao, Sheng, Zhu, Chan,
  Mou, Luo, Xiang, Zhou, Zhu, and Deng}]{can}
Bian, W.; Wu, K.; Ren, L.; Pi, Q.; Zhang, Y.; Xiao, C.; Sheng, X.; Zhu, Y.;
  Chan, Z.; Mou, N.; Luo, X.; Xiang, S.; Zhou, G.; Zhu, X.; and Deng, H. 2022.
\newblock {CAN:} Feature Co-Action Network for Click-Through Rate Prediction.
\newblock In Candan, K.~S.; Liu, H.; Akoglu, L.; Dong, X.~L.; and Tang, J.,
  eds., \emph{{WSDM} '22: The Fifteenth {ACM} International Conference on Web
  Search and Data Mining, Virtual Event / Tempe, AZ, USA, February 21 - 25,
  2022}, 57--65. {ACM}.

\bibitem[{Bian et~al.(2021)Bian, Zhou, Fu, Yang, Sun, Tang, Liu, Liu, and
  Li}]{dumn}
Bian, Z.; Zhou, S.; Fu, H.; Yang, Q.; Sun, Z.; Tang, J.; Liu, G.; Liu, K.; and
  Li, X. 2021.
\newblock Denoising User-aware Memory Network for Recommendation.
\newblock In Pamp{\'{\i}}n, H. J.~C.; Larson, M.~A.; Willemsen, M.~C.; Konstan,
  J.~A.; McAuley, J.~J.; Garcia{-}Gathright, J.; Huurnink, B.; and Oldridge,
  E., eds., \emph{RecSys '21: Fifteenth {ACM} Conference on Recommender
  Systems, Amsterdam, The Netherlands, 27 September 2021 - 1 October 2021},
  400--410. {ACM}.

\bibitem[{Chang et~al.(2010)Chang, Hsieh, Chang, Ringgaard, and Lin}]{poly2}
Chang, Y.; Hsieh, C.; Chang, K.; Ringgaard, M.; and Lin, C. 2010.
\newblock Training and Testing Low-degree Polynomial Data Mappings via Linear
  {SVM}.
\newblock \emph{J. Mach. Learn. Res.}, 11: 1471--1490.

\bibitem[{Chen et~al.(2018{\natexlab{a}})Chen, Liu, Zha, Zhou, Xiong, and
  Li}]{thacil}
Chen, X.; Liu, D.; Zha, Z.; Zhou, W.; Xiong, Z.; and Li, Y. 2018{\natexlab{a}}.
\newblock Temporal Hierarchical Attention at Category- and Item-Level for
  Micro-Video Click-Through Prediction.
\newblock In Boll, S.; Lee, K.~M.; Luo, J.; Zhu, W.; Byun, H.; Chen, C.~W.;
  Lienhart, R.; and Mei, T., eds., \emph{2018 {ACM} Multimedia Conference on
  Multimedia Conference, {MM} 2018, Seoul, Republic of Korea, October 22-26,
  2018}, 1146--1153. {ACM}.

\bibitem[{Chen et~al.(2018{\natexlab{b}})Chen, Xu, Zhang, Tang, Cao, Qin, and
  Zha}]{rum}
Chen, X.; Xu, H.; Zhang, Y.; Tang, J.; Cao, Y.; Qin, Z.; and Zha, H.
  2018{\natexlab{b}}.
\newblock Sequential Recommendation with User Memory Networks.
\newblock In Chang, Y.; Zhai, C.; Liu, Y.; and Maarek, Y., eds.,
  \emph{Proceedings of the Eleventh {ACM} International Conference on Web
  Search and Data Mining, {WSDM} 2018, Marina Del Rey, CA, USA, February 5-9,
  2018}, 108--116. {ACM}.

\bibitem[{Cheng et~al.(2016)Cheng, Koc, Harmsen, Shaked, Chandra, Aradhye,
  Anderson, Corrado, Chai, Ispir, Anil, Haque, Hong, Jain, Liu, and
  Shah}]{cheng2016wide}
Cheng, H.; Koc, L.; Harmsen, J.; Shaked, T.; Chandra, T.; Aradhye, H.;
  Anderson, G.; Corrado, G.; Chai, W.; Ispir, M.; Anil, R.; Haque, Z.; Hong,
  L.; Jain, V.; Liu, X.; and Shah, H. 2016.
\newblock Wide {\&} Deep Learning for Recommender Systems.
\newblock In Karatzoglou, A.; Hidasi, B.; Tikk, D.; Shalom, O.~S.; Roitman, H.;
  Shapira, B.; and Rokach, L., eds., \emph{Proceedings of the 1st Workshop on
  Deep Learning for Recommender Systems, DLRS@RecSys 2016, Boston, MA, USA,
  September 15, 2016}, 7--10. {ACM}.

\bibitem[{Chung et~al.(2014)Chung, G{\"{u}}l{\c{c}}ehre, Cho, and Bengio}]{gru}
Chung, J.; G{\"{u}}l{\c{c}}ehre, {\c{C}}.; Cho, K.; and Bengio, Y. 2014.
\newblock Empirical Evaluation of Gated Recurrent Neural Networks on Sequence
  Modeling.
\newblock \emph{CoRR}, abs/1412.3555.

\bibitem[{Graves, Wayne, and Danihelka(2014)}]{ntm}
Graves, A.; Wayne, G.; and Danihelka, I. 2014.
\newblock Neural Turing Machines.
\newblock \emph{CoRR}, abs/1410.5401.

\bibitem[{Guo et~al.(2017)Guo, Tang, Ye, Li, and He}]{guo_deepfm}
Guo, H.; Tang, R.; Ye, Y.; Li, Z.; and He, X. 2017.
\newblock DeepFM: {A} Factorization-Machine based Neural Network for {CTR}
  Prediction.
\newblock In Sierra, C., ed., \emph{Proceedings of the Twenty-Sixth
  International Joint Conference on Artificial Intelligence, {IJCAI} 2017,
  Melbourne, Australia, August 19-25, 2017}, 1725--1731. ijcai.org.

\bibitem[{He et~al.(2016)He, Zhang, Ren, and Sun}]{resnet}
He, K.; Zhang, X.; Ren, S.; and Sun, J. 2016.
\newblock Deep Residual Learning for Image Recognition.
\newblock In \emph{2016 {IEEE} Conference on Computer Vision and Pattern
  Recognition, {CVPR} 2016, Las Vegas, NV, USA, June 27-30, 2016}, 770--778.
  {IEEE} Computer Society.

\bibitem[{Hidasi et~al.(2016)Hidasi, Karatzoglou, Baltrunas, and
  Tikk}]{gru4rec}
Hidasi, B.; Karatzoglou, A.; Baltrunas, L.; and Tikk, D. 2016.
\newblock Session-based Recommendations with Recurrent Neural Networks.
\newblock In Bengio, Y.; and LeCun, Y., eds., \emph{4th International
  Conference on Learning Representations, {ICLR} 2016, San Juan, Puerto Rico,
  May 2-4, 2016, Conference Track Proceedings}.

\bibitem[{Hochreiter and Schmidhuber(1997)}]{lstm}
Hochreiter, S.; and Schmidhuber, J. 1997.
\newblock Long Short-Term Memory.
\newblock \emph{Neural Comput.}, 9(8): 1735--1780.

\bibitem[{Kang and McAuley(2018)}]{sasrec}
Kang, W.; and McAuley, J.~J. 2018.
\newblock Self-Attentive Sequential Recommendation.
\newblock In \emph{{IEEE} International Conference on Data Mining, {ICDM} 2018,
  Singapore, November 17-20, 2018}, 197--206. {IEEE} Computer Society.

\bibitem[{Kingma and Ba(2015)}]{adam}
Kingma, D.~P.; and Ba, J. 2015.
\newblock Adam: {A} Method for Stochastic Optimization.
\newblock In Bengio, Y.; and LeCun, Y., eds., \emph{3rd International
  Conference on Learning Representations, {ICLR} 2015, San Diego, CA, USA, May
  7-9, 2015, Conference Track Proceedings}.

\bibitem[{Lian et~al.(2018)Lian, Zhou, Zhang, Chen, Xie, and Sun}]{xdeepfm}
Lian, J.; Zhou, X.; Zhang, F.; Chen, Z.; Xie, X.; and Sun, G. 2018.
\newblock xDeepFM: Combining Explicit and Implicit Feature Interactions for
  Recommender Systems.
\newblock In Guo, Y.; and Farooq, F., eds., \emph{Proceedings of the 24th {ACM}
  {SIGKDD} International Conference on Knowledge Discovery {\&} Data Mining,
  {KDD} 2018, London, UK, August 19-23, 2018}, 1754--1763. {ACM}.

\bibitem[{Lv et~al.(2020)Lv, Li, Guo, Yu, Sun, Jin, and Yang}]{sru2b}
Lv, F.; Li, M.; Guo, T.; Yu, C.; Sun, F.; Jin, T.; and Yang, K. 2020.
\newblock Unclicked User Behaviors Enhanced Sequential Recommendation.
\newblock \emph{CoRR}, abs/2010.12837.

\bibitem[{McMahan et~al.(2013)McMahan, Holt, Sculley, Young, Ebner, Grady, Nie,
  Phillips, Davydov, Golovin, Chikkerur, Liu, Wattenberg, Hrafnkelsson, Boulos,
  and Kubica}]{ftrl}
McMahan, H.~B.; Holt, G.; Sculley, D.; Young, M.; Ebner, D.; Grady, J.; Nie,
  L.; Phillips, T.; Davydov, E.; Golovin, D.; Chikkerur, S.; Liu, D.;
  Wattenberg, M.; Hrafnkelsson, A.~M.; Boulos, T.; and Kubica, J. 2013.
\newblock Ad click prediction: a view from the trenches.
\newblock In Dhillon, I.~S.; Koren, Y.; Ghani, R.; Senator, T.~E.; Bradley, P.;
  Parekh, R.; He, J.; Grossman, R.~L.; and Uthurusamy, R., eds., \emph{The 19th
  {ACM} {SIGKDD} International Conference on Knowledge Discovery and Data
  Mining, {KDD} 2013, Chicago, IL, USA, August 11-14, 2013}, 1222--1230. {ACM}.

\bibitem[{Meng et~al.(2023)Meng, Zhang, Guo, Guo, Liu, Zhang, Zheng, Tang, Li,
  and Zhang}]{hpmr}
Meng, C.; Zhang, H.; Guo, W.; Guo, H.; Liu, H.; Zhang, Y.; Zheng, H.; Tang, R.;
  Li, X.; and Zhang, R. 2023.
\newblock Hierarchical Projection Enhanced Multi-behavior Recommendation.
\newblock In Singh, A.; Sun, Y.; Akoglu, L.; Gunopulos, D.; Yan, X.; Kumar, R.;
  Ozcan, F.; and Ye, J., eds., \emph{Proceedings of the 29th {ACM} {SIGKDD}
  Conference on Knowledge Discovery and Data Mining, {KDD} 2023, Long Beach,
  CA, USA, August 6-10, 2023}, 4649--4660. {ACM}.

\bibitem[{Ouyang et~al.(2019)Ouyang, Zhang, Li, Zou, Xing, Liu, and Du}]{dstn}
Ouyang, W.; Zhang, X.; Li, L.; Zou, H.; Xing, X.; Liu, Z.; and Du, Y. 2019.
\newblock Deep Spatio-Temporal Neural Networks for Click-Through Rate
  Prediction.
\newblock In Teredesai, A.; Kumar, V.; Li, Y.; Rosales, R.; Terzi, E.; and
  Karypis, G., eds., \emph{Proceedings of the 25th {ACM} {SIGKDD} International
  Conference on Knowledge Discovery {\&} Data Mining, {KDD} 2019, Anchorage,
  AK, USA, August 4-8, 2019}, 2078--2086. {ACM}.

\bibitem[{Pi et~al.(2019)Pi, Bian, Zhou, Zhu, and Gai}]{mimn}
Pi, Q.; Bian, W.; Zhou, G.; Zhu, X.; and Gai, K. 2019.
\newblock Practice on Long Sequential User Behavior Modeling for Click-Through
  Rate Prediction.
\newblock In Teredesai, A.; Kumar, V.; Li, Y.; Rosales, R.; Terzi, E.; and
  Karypis, G., eds., \emph{Proceedings of the 25th {ACM} {SIGKDD} International
  Conference on Knowledge Discovery {\&} Data Mining, {KDD} 2019, Anchorage,
  AK, USA, August 4-8, 2019}, 2671--2679. {ACM}.

\bibitem[{Qu et~al.(2016)Qu, Cai, Ren, Zhang, Yu, Wen, and Wang}]{pnn}
Qu, Y.; Cai, H.; Ren, K.; Zhang, W.; Yu, Y.; Wen, Y.; and Wang, J. 2016.
\newblock Product-Based Neural Networks for User Response Prediction.
\newblock In Bonchi, F.; Domingo{-}Ferrer, J.; Baeza{-}Yates, R.; Zhou, Z.; and
  Wu, X., eds., \emph{{IEEE} 16th International Conference on Data Mining,
  {ICDM} 2016, December 12-15, 2016, Barcelona, Spain}, 1149--1154. {IEEE}
  Computer Society.

\bibitem[{Rendle(2010)}]{fm}
Rendle, S. 2010.
\newblock Factorization Machines.
\newblock In Webb, G.~I.; Liu, B.; Zhang, C.; Gunopulos, D.; and Wu, X., eds.,
  \emph{{ICDM} 2010, The 10th {IEEE} International Conference on Data Mining,
  Sydney, Australia, 14-17 December 2010}, 995--1000. {IEEE} Computer Society.

\bibitem[{Rendle(2012)}]{libfm}
Rendle, S. 2012.
\newblock Factorization Machines with libFM.
\newblock \emph{{ACM} Trans. Intell. Syst. Technol.}, 3(3): 57:1--57:22.

\bibitem[{Simonyan and Zisserman(2015)}]{vgg}
Simonyan, K.; and Zisserman, A. 2015.
\newblock Very Deep Convolutional Networks for Large-Scale Image Recognition.
\newblock In Bengio, Y.; and LeCun, Y., eds., \emph{3rd International
  Conference on Learning Representations, {ICLR} 2015, San Diego, CA, USA, May
  7-9, 2015, Conference Track Proceedings}.

\bibitem[{Sun et~al.(2019)Sun, Liu, Wu, Pei, Lin, Ou, and Jiang}]{bert4rec}
Sun, F.; Liu, J.; Wu, J.; Pei, C.; Lin, X.; Ou, W.; and Jiang, P. 2019.
\newblock BERT4Rec: Sequential Recommendation with Bidirectional Encoder
  Representations from Transformer.
\newblock In Zhu, W.; Tao, D.; Cheng, X.; Cui, P.; Rundensteiner, E.~A.;
  Carmel, D.; He, Q.; and Yu, J.~X., eds., \emph{Proceedings of the 28th {ACM}
  International Conference on Information and Knowledge Management, {CIKM}
  2019, Beijing, China, November 3-7, 2019}, 1441--1450. {ACM}.

\bibitem[{Tang and Wang(2018)}]{caser}
Tang, J.; and Wang, K. 2018.
\newblock Personalized Top-N Sequential Recommendation via Convolutional
  Sequence Embedding.
\newblock In Chang, Y.; Zhai, C.; Liu, Y.; and Maarek, Y., eds.,
  \emph{Proceedings of the Eleventh {ACM} International Conference on Web
  Search and Data Mining, {WSDM} 2018, Marina Del Rey, CA, USA, February 5-9,
  2018}, 565--573. {ACM}.

\bibitem[{Vaswani et~al.(2017)Vaswani, Shazeer, Parmar, Uszkoreit, Jones,
  Gomez, Kaiser, and Polosukhin}]{transformer}
Vaswani, A.; Shazeer, N.; Parmar, N.; Uszkoreit, J.; Jones, L.; Gomez, A.~N.;
  Kaiser, L.; and Polosukhin, I. 2017.
\newblock Attention is All you Need.
\newblock In Guyon, I.; von Luxburg, U.; Bengio, S.; Wallach, H.~M.; Fergus,
  R.; Vishwanathan, S. V.~N.; and Garnett, R., eds., \emph{Advances in Neural
  Information Processing Systems 30: Annual Conference on Neural Information
  Processing Systems 2017, December 4-9, 2017, Long Beach, CA, {USA}},
  5998--6008.

\bibitem[{Wang et~al.(2017)Wang, Fu, Fu, and Wang}]{dcn}
Wang, R.; Fu, B.; Fu, G.; and Wang, M. 2017.
\newblock Deep {\&} Cross Network for Ad Click Predictions.
\newblock In \emph{Proceedings of the ADKDD'17, Halifax, NS, Canada, August 13
  - 17, 2017}, 12:1--12:7. {ACM}.

\bibitem[{Xie et~al.(2020)Xie, Ling, Wang, Wang, Xia, and Lin}]{dfn}
Xie, R.; Ling, C.; Wang, Y.; Wang, R.; Xia, F.; and Lin, L. 2020.
\newblock Deep Feedback Network for Recommendation.
\newblock In Bessiere, C., ed., \emph{Proceedings of the Twenty-Ninth
  International Joint Conference on Artificial Intelligence, {IJCAI} 2020},
  2519--2525. ijcai.org.

\bibitem[{Zhou et~al.(2019)Zhou, Mou, Fan, Pi, Bian, Zhou, Zhu, and Gai}]{dien}
Zhou, G.; Mou, N.; Fan, Y.; Pi, Q.; Bian, W.; Zhou, C.; Zhu, X.; and Gai, K.
  2019.
\newblock Deep Interest Evolution Network for Click-Through Rate Prediction.
\newblock In \emph{The Thirty-Third {AAAI} Conference on Artificial
  Intelligence, {AAAI} 2019, The Thirty-First Innovative Applications of
  Artificial Intelligence Conference, {IAAI} 2019, The Ninth {AAAI} Symposium
  on Educational Advances in Artificial Intelligence, {EAAI} 2019, Honolulu,
  Hawaii, USA, January 27 - February 1, 2019}, 5941--5948. {AAAI} Press.

\bibitem[{Zhou et~al.(2018)Zhou, Zhu, Song, Fan, Zhu, Ma, Yan, Jin, Li, and
  Gai}]{din}
Zhou, G.; Zhu, X.; Song, C.; Fan, Y.; Zhu, H.; Ma, X.; Yan, Y.; Jin, J.; Li,
  H.; and Gai, K. 2018.
\newblock Deep Interest Network for Click-Through Rate Prediction.
\newblock In Guo, Y.; and Farooq, F., eds., \emph{Proceedings of the 24th {ACM}
  {SIGKDD} International Conference on Knowledge Discovery {\&} Data Mining,
  {KDD} 2018, London, UK, August 19-23, 2018}, 1059--1068. {ACM}.

\end{thebibliography}
\newpage
\appendix
\setcounter{secnumdepth}{1}
\renewcommand{\thetable}{\Alph{section}\arabic{table}}
\renewcommand{\thefigure}{\Alph{section}\arabic{figure}}

\section{Notation Definition}

For ease of presentation, we provide a summary of the main notations used in this paper in Table \ref{tab:notation}.

\begin{table}[!ht]
 \resizebox{\columnwidth}{!}{%
\begin{tabular}{cl}
\toprule
Notation  &  Description \\ \hline
$u$ & The target user.    \\ 
$v_{tgt}$ & The target item. \\
$\theta$ & The parameters of the CTR prediction model.\\
$n$ & The maximum length of the user click sequence.\\
$l$ & The range of exposure context.\\
$\mathcal{S}$ & The displayed sequence of items.\\
$\mathbf{o}^c_j$ & The one-hot vector of category feature.\\
\multirow{2}{*}{$\mathbf{o}^f_j$} & The high-dimensional feature vector in micro-\\ &video dataset.\\
$\mathbf{E}^c, \mathbf{E}^f$ & The linear-mapping embedding matrix.\\
$\mathcal{S}_{clk}, \mathcal{S}_{unclk}$ & The sequences of clicked and unclicked items. \\
$v_j^{clk}, v_j^{unclk}$  & The $j$-th clicked and unclicked item. \\
\multirow{2}{*}{$S(v^{clk}_j)$} & The set of exposed but unclicked items retrieved\\ & by $v_j^{clk}$.\\
\multirow{3}{*}{$\mathbf{e}^c_j, \mathbf{e}^f_j$} & Embedding of category features and embedding\\ & extracted from the cover images in the micro video\\ & scene. \\
\multirow{2}{*}{$\mathcal{S}_{clk}^{emb}, \mathcal{S}_{unclk}^{emb}$} & The corresponding embedding sequence of $\mathcal{S}_{clk},$\\ & $\mathcal{S}_{unclk}$.\\
$\mathcal{S}(v^{clk}_j)^{emb}$ & The corresponding embedding sequence of $S(v^{clk}_j)$.\\
\multirow{2}{*}{$\mathbf{c}_j$} & Exposure contextual information representation \\ &of $v_j^{clk}$. \\
$\mathbf{e}_j^{clk*}$ & The
enhanced click behavior representation of $v_j^{clk}$. \\
$\mathbf{e}_{tgt}$ & The target item embedding $v_{tgt}$. \\
\multirow{2}{*}{$\mathbf{h}$} & Representation extracted from $\mathbf{e}_j^{clk*}$ that are guided\\ & by $\mathbf{e}_{tgt}$.\\
\multirow{2}{*}{$\mathbf{g}$} & Representation extracted from $\mathbf{c}_j$ that are guided by\\ & $\mathbf{e}_{tgt}$.\\
\multirow{2}{*}{$y_i, p_i$} & The label and the predicted probability of the $i$-th\\ & user-item interaction. \\
\bottomrule
\end{tabular}}
\caption{Main notations and corresponding descriptions}
\label{tab:notation}
\end{table}

\section{Baseline Methods}

To comprehensively demonstrate the effectiveness of the proposed TEM4CTR, we compared it with nine state-of-the-art recommendation baselines: 

\begin{itemize}
\item \textbf{AvgPool DNN} is the basic deep learning model for CTR
prediction. It incorporates behavior embeddings via sum-pooling operation.
\item \textbf{RUM} \cite{rum} leverages an external memory to store user behavior features and employs soft-writing and attention-reading mechanisms for interaction with memory. Feature-level RUM is applied as a baseline to extract user behavior sequence information in our experiment.
\item \textbf{ARNN} \cite{mimn} is a variation of "GRU4Rec" that employs an attention mechanism for weighted summing of hidden states along time to improve user sequence representation.
\item \textbf{GRU4Rec} \cite{gru4rec} uses the GRU to model user behavior sequences and is the first work using the recurrent cell for user behavior sequence modeling.
\item \textbf{DIN} \cite{din} is based on a target attention mechanism to dynamically capture the latent interest of users in a given target item.
\item \textbf{DIEN} \cite{dien} makes improvements to DIN. It introduces a GRU network to model the temporal evolution of user interests and designs auxiliary losses to help the extraction of users' latent interest.
\item \textbf{DSTN} \cite{dstn} investigates various types of auxiliary context for improving the CTR prediction of the target ad.
\item \textbf{DFN} \cite{dfn} considers multiple explicit and implicit feedback information and the connections between them, thus improving the recommendation effectiveness through multi-feedback modeling. 
\item \textbf{CAN} \cite{can} decouples the feature interaction modeling from the original feature modeling by mapping the features into a mini-MLP to interact with other features. For a fair comparison, DIN was used as the base model for CAN.
\end{itemize}

\section{RQ5: Hyperparameter Analysis}

\begin{figure}[!ht]
    \centering
    \begin{subfigure}{0.23\textwidth}
        \centering
        \includegraphics[width=1.65in]{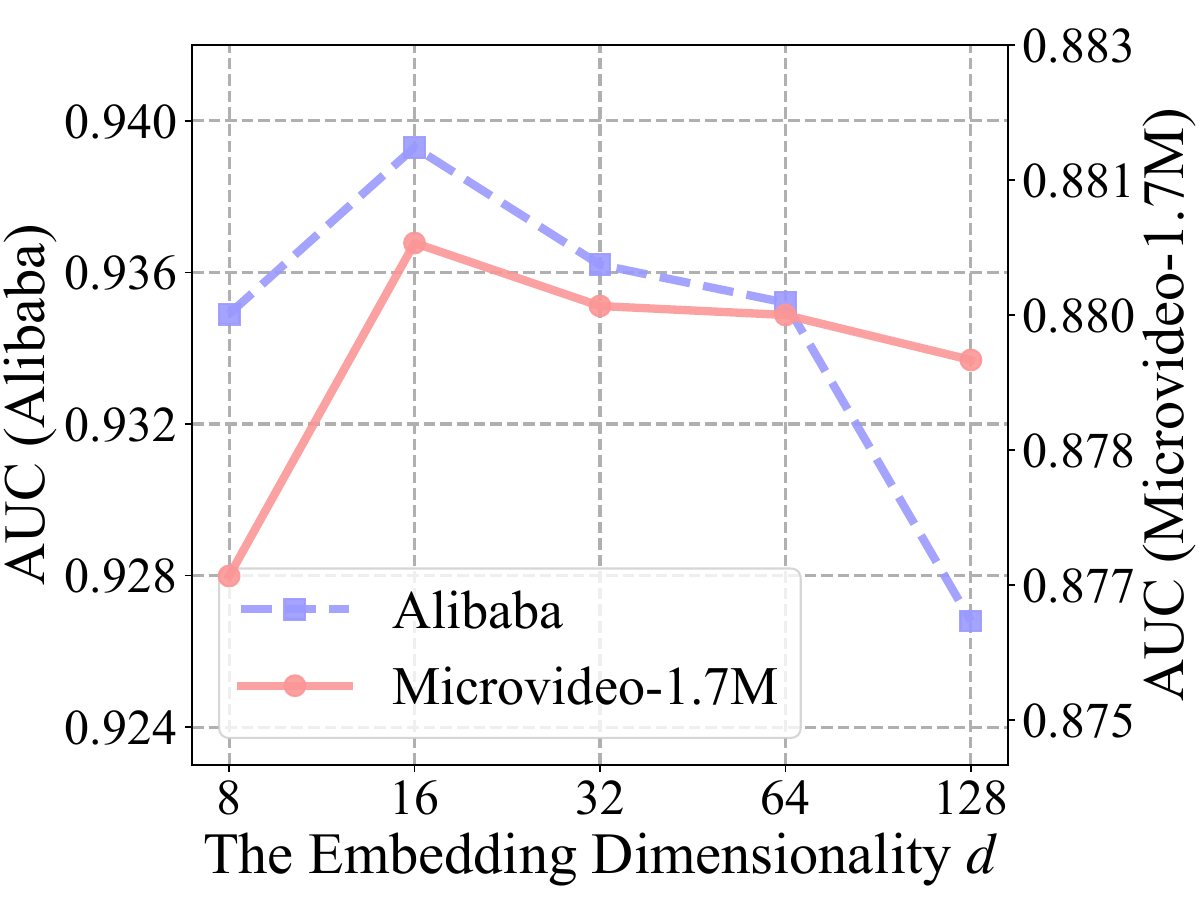} 
        \caption{Embedding dimensionality $d$}
        \label{fig:hyper1}
    \end{subfigure}
    \begin{subfigure}{0.23\textwidth}
        \centering
        \includegraphics[width=1.65in]{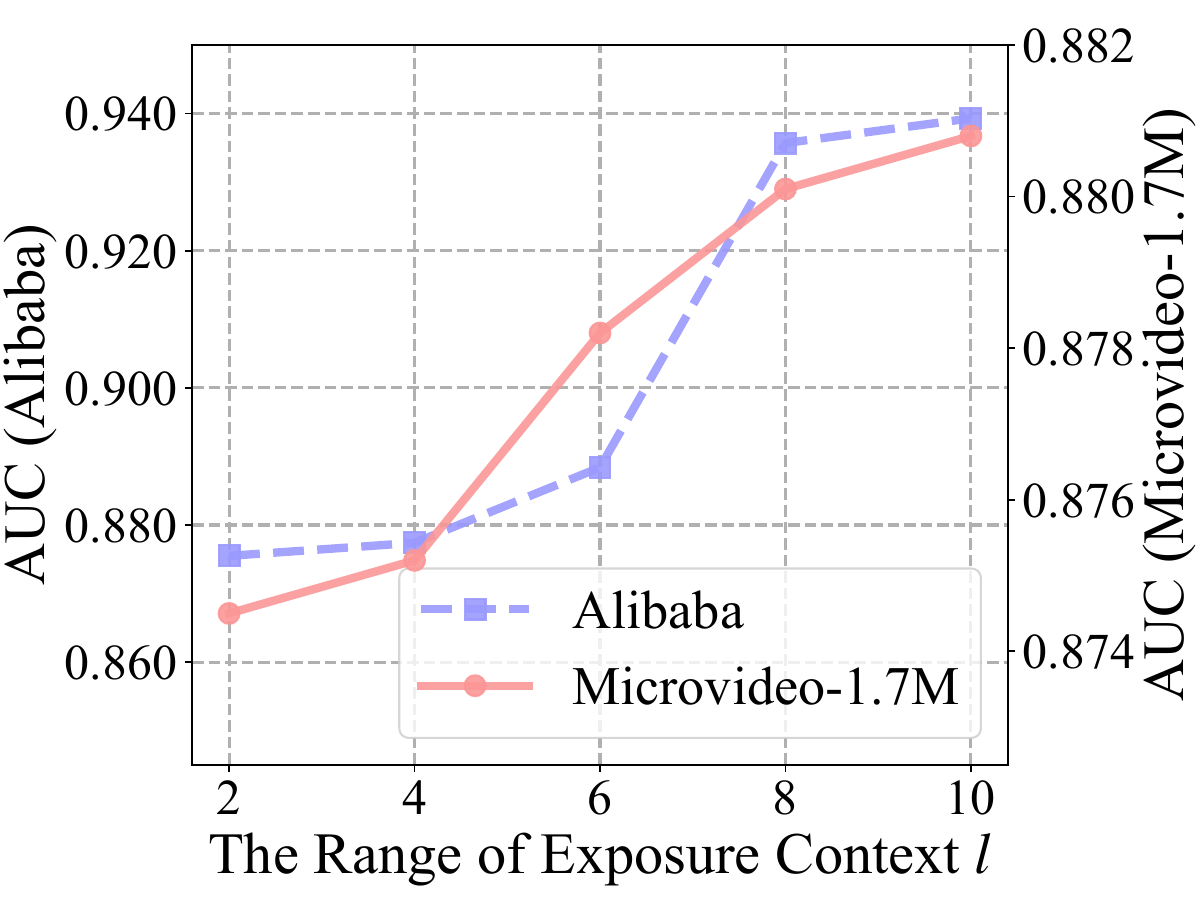} 
        \caption{Exposure context range $l$}
        \label{fig:hyper2}
    \end{subfigure}
        \caption{Performances with different hyperparameter settings.}
    
    \end{figure}

To investigate the effect of different hyperparameters on TEM4CTR, we conduct experiments on two datasets with different settings of the key hyperparameters, including (1) the embedding dimensionality $d$, and (2) the range of exposure context $l$.
When studying the hyperparameters of interest, we keep all other hyperparameters fixed.

\subsubsection{Impact of the embedding dimensionality $d$.}
The embedding dimensionality $d$ determines the model's capacity and thus further affects the learning of representations. As shown in Figure \ref{fig:hyper1}, we adjusted $d$ from $\{8,16,32,64,128\}$ on Alibaba and  Microvideo datasets respectively, and the best performance can be achieved with the embedding dimensionality of 16 on both of the two datasets. With the increase of embedding dimensionality, the performance of the model declines on both datasets. One possible reason is the overfitting issue with a larger number of hidden representation spaces. Meanwhile, the performance degradation is more significant on Alibaba than on Microvideo, which may be due to the additional picture pre-training representations in micro video, which provided more semantic information for model learning, thus alleviating the overfitting.

\subsubsection{Impact of the range of exposure context $l$.}
The range of exposure context is the length of the unclicked sequence searched by the clicked item. It determines the amount of data used for exposure context information extraction. As shown in Figure \ref{fig:hyper2}, we adjust the range of exposure context in the scope $\{2,4,6,8,10\}$. It can be seen that with the increase of $l$, the performance of the model continues to improve, and the improvement amplitude is approximately "S" shaped curve. At shorter $l$, the performance growth is slow and limited, which may be because the context range is relatively short, and the searched unclicked feedback is not sufficient to well reflect the exposure context information corresponding to the click behavior. With $l$ increasing further, the model performance is rapidly improved. When $l$ continues to increase, we will find that the improvement effect of the model is limited and slow, and the marginal effect is diminishing. This may be because the exposure context information is sufficient to model the user's latent interest in this setting, so it is not cost-effective to continue to increase the range of the exposure context.

\end{document}